\documentclass[iop,apj]{emulateapj}
\usepackage{amsmath}
\def\msun{M$_\odot$}

\def\lsim{\mathrel{\rlap{\lower4pt\hbox{\hskip1pt$\sim$}}    \raise1pt\hbox{$<$}}} 
\def\aj{AJ}
\def\apj{ApJ}
\def\apjs{ApJS}
\def\apjl{ApJL}
\def\mnras{MNRAS}
\def\aap{A\&A}

\def\pasp{PASP}

\def\g{$g_{475}$}
\def\V{$V_{606}$}
\def\I{$I_{814}$}
\def\z{$z_{850}$}

\shorttitle{The dwarf nature of ultra diffuse galaxies}
\shortauthors{Beasley \& Trujillo}

\begin{document}

\title{Globular clusters indicate ultra diffuse galaxies are dwarfs}

\author{Michael A.\ Beasley\altaffilmark{1,2} and 
Ignacio Trujillo\altaffilmark{1,2}}
  
\email{beasley@iac.es}

\altaffiltext{1}{Instituto de Astrof\'isica de Canarias, Calle Via L\'actea, La Laguna, Tenerife, Spain}
\altaffiltext{2}{University of La Laguna. Avda. Astrof\'isico Fco. S\'anchez, La Laguna, Tenerife, Spain}

\begin{abstract}

  We present an analysis of archival {\it HST/ACS} imaging in the F475W (\g), F606W (\V) and F814W (\I) bands of the
  globular cluster (GC) system of a large (3.4 kpc effective radius) ultra-diffuse galaxy (DF17) believed located in
  the Coma Cluster of galaxies. We detect 11 GCs down to the 5$\sigma$ completeness limit of the imaging (\I=27 mag).
  Correcting for background and our detection limits yields a total population of GCs in this galaxy of  $27\pm5$
  and a $V$-band specific frequency,  $S_N=28\pm5$.
  Based on comparisons to the GC systems of Local galaxies, we show that both the absolute number and
  the colors of the GC system of DF17 are consistent with the
  GC system of a dark-matter dominated dwarf galaxy with virial mass $\sim9.0\times10^{10}$~\msun~ and a
  dark-to-stellar mass ratio, $M_{vir} / M_{ star}\sim 1000$.
  Based on the stellar mass-growth of the
  Milky Way, we show that DF17  cannot be understood as a failed Milky Way-like system, but is more
  similar to quenched Large Magellanic Cloud-like systems. 
  We find that the mean color of GC population, \g--\I =  $0.91\pm0.05$ mag, 
  coincides with the peak of the color distribution of intracluster GCs and are also
  similar to those of the blue GCs in the outer regions of massive galaxies. We suggest that both the intracluster GC
  population in Coma and the blue-peak in the GC populations of massive galaxies may be fed - at least in part - by the
  disrupted equivalents of systems such as DF17.
 \end{abstract}


\section{Introduction}

Ultra-diffuse galaxies (UDGs - a termed coined by van Dokkum et al. 2015) constitute a likely heterogeneous
population of low-surface brightness systems (e.g. Impey
et al. 1988; Bothun et al. 1991; Dalcanton et al. 1997; van Dokkum et al. 2015; Koda et al. 2015; Mihos et al. 2015;
Mu\~noz et al. 2015; van der Burg et al. 2016; Martinez-Delgado et al. 2016). Their most outstanding
characteristic is that while their stellar masses
are of order $10^8$ \msun~ or less, their effective radii are comparable to L$_\star$ galaxies (2 - 5 kpc). For this
reason, the origin of the UDGs remains controversial. One possibility discussed by van Dokkum et al. (2015)
is that UDGs are failed L$_\star$ galaxies whose early accretion to the cluster
environment (z$\sim$2) quenched their growth. Others, however, have suggested that UDGs are the high-spin
tail of normal dwarf galaxies (Amorisco \& Loeb 2016).

The two competing scenarios can be confronted with the observations using different approaches. A strong
test is to measure the virial mass of UDGs and explore whether they inhabit
dark matter (DM) haloes of dwarf-mass ($M_{vir}\sim 10^{10-11}$ \msun) or Milky Way (MW)-mass haloes ($M_{vir}\sim 10^{12}$ \msun).
UDGs are generally too faint to measure dynamical masses directly through stellar velocity dispersions and rotation curves.
Beasley et al. (2016) have shown that the dynamics of the globular cluster (GC) systems  of UDGs in the Virgo cluster can be
used to measure dynamical masses. These authors determined a total virial mass of $(8\pm4)\times 10^{10}$ \msun~for the
UDG VCC~1287, in agreement with the hypothesis that UDGs are dwarf-like systems.
An alternative, followed by
Rom\'an \& Trujillo (2016), is to explore the spatial distribution of UDGs versus dwarfs and L$_\star$ galaxies in clusters and
outside these structures. These authors also find that the spatial distribution of UDGs are more similar to dwarfs
than to L$_\star$ galaxies. 

In this contribution we determine the total number of GCs in a large
UDG (DF17; van Dokkum et al. 2015) in the Coma cluster based on archival {\it Hubble Space Telescope} imaging.
This galaxy has a size (effective radius 3.4 kpc) large enough to be considered as a good candidate for a
failed MW galaxy at high redshift, and is the fourth most luminous UDG in the van Dokkum et al. (2015) catalog.
If this UDG were an early quenched L$_\star$ galaxy, a natural expectation would be
that the number of GCs in DF17, and the GC mean colors,  will be  similar to the in-situ GCs of the MW.
As we will show in this work, the total number of GCs in DF17, and the mean GC colors, imply that they are hosted by
a dwarf rather than a giant galaxy. 

During the refereeing process of this paper, Peng \& Lim (2016) published a study based on the same
  dataset analysed here. These authors obtain very similar results for the total number of GCs
  in DF17 ($N_{\rm GC}=28\pm14$ GCs), but focus on slightly different aspects of the GC system, namely the GC luminosity function
and density profile rather than color distributions.

\section{Data}

\subsection{Image preparation}

\begin{figure*}[ht]
\epsscale{1.1}
\plotone{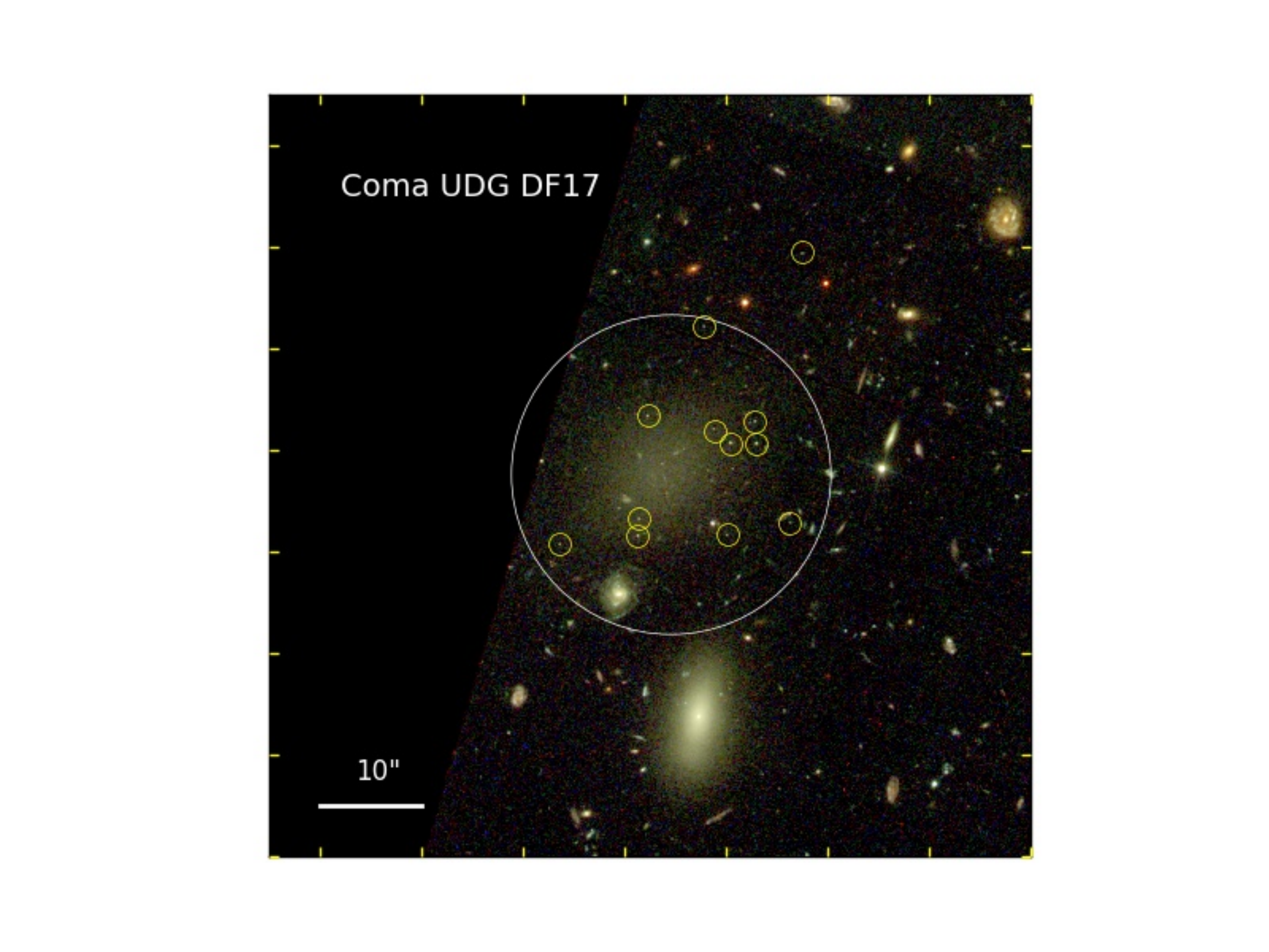}
\caption{\g,\V,\I~ {\it HST/ACS} composite image centered on DF17. GC candidates are marked with
  small yellow circles. The large white circle represents a 300 pixel radius region (15$''$,
  or $\sim2$ times the galaxy effective radius) within which we associate GC candidates as those
  belonging to DF17. 
}\label{fig:Fig1}
\end{figure*}

van Dokkum et al. (2015) identified a sample of 47 UDGs projected
against the Coma cluster with the Dragonfly telephoto array (Abraham \& van Dokkum 2015).
One of these UDGs, DF17, van Dokkum et al. re-identified in archival {\it HST/ACS} imaging.

We retrieved these data in order to analyse the GC population of this system.
These data comprise of deep imaging in the F475W (\g), F606W (\V) and
F814W (\I) filters (GO-12476; PI: Cook; Macri et al. 2013).
Exposure times totalled 5100s, 5820s and 5100s in each filter, respectively.

These data were retrieved from the archive on 18 March 2016  and after pipeline reduction,
individual exposures were median combined with SWARP (Bertin et al. 2002) in order to place them
on a common grid  and remove a significant cosmic ray contribution.
The combined 3-band color image zooming in the the region around DF17 is shown in Fig.~\ref{fig:Fig1}.

\subsection{Globular cluster photometry and selection}

We performed aperture photometry on the imaging data
using SExtractor (Bertin 1996) with a 5-pixel radial
aperture. In order to maximize our object
detection, we set both the detection and analysis thresholds
to 0.9$\sigma$ of the background level, and detected objects
on 7-pixel unsharp masked images in the F814W filter.
We required that objects were detected
within all three filters to be considered a real source. 
We performed photometry on the original images.
Magnitude zeropoints were obtained from the ACS webpages corresponding
to the dates of the observations, and are ZP(F475W) = 26.06,
ZP(F606W) = 26.49 and ZP(F814W) = 25.94. AB magnitudes were
then obtained by correcting to infinite aperture using the enclosed
energy curves of Sirianni et al. (2005).
For the above radial aperture, we determine that the {\it HST/ACS} imaging
is 5$\sigma$ complete down to F475W = 27.20, F606W = 27.33 and F814W = 26.98.

At the distance of the Coma cluster ($m-M$=35.0; Carter et al. 2008),
GCs are unresolved and appear as point-sources. However, the majority of background galaxies are
resolved allowing for their effective removal.
We selected point sources by running our photometry with two aperture sizes, one of 5 pixel
radius and another of 10 pixels similar to the approach of Peng et al. (2011).

\begin{figure*}
\epsscale{1.1}
\plotone{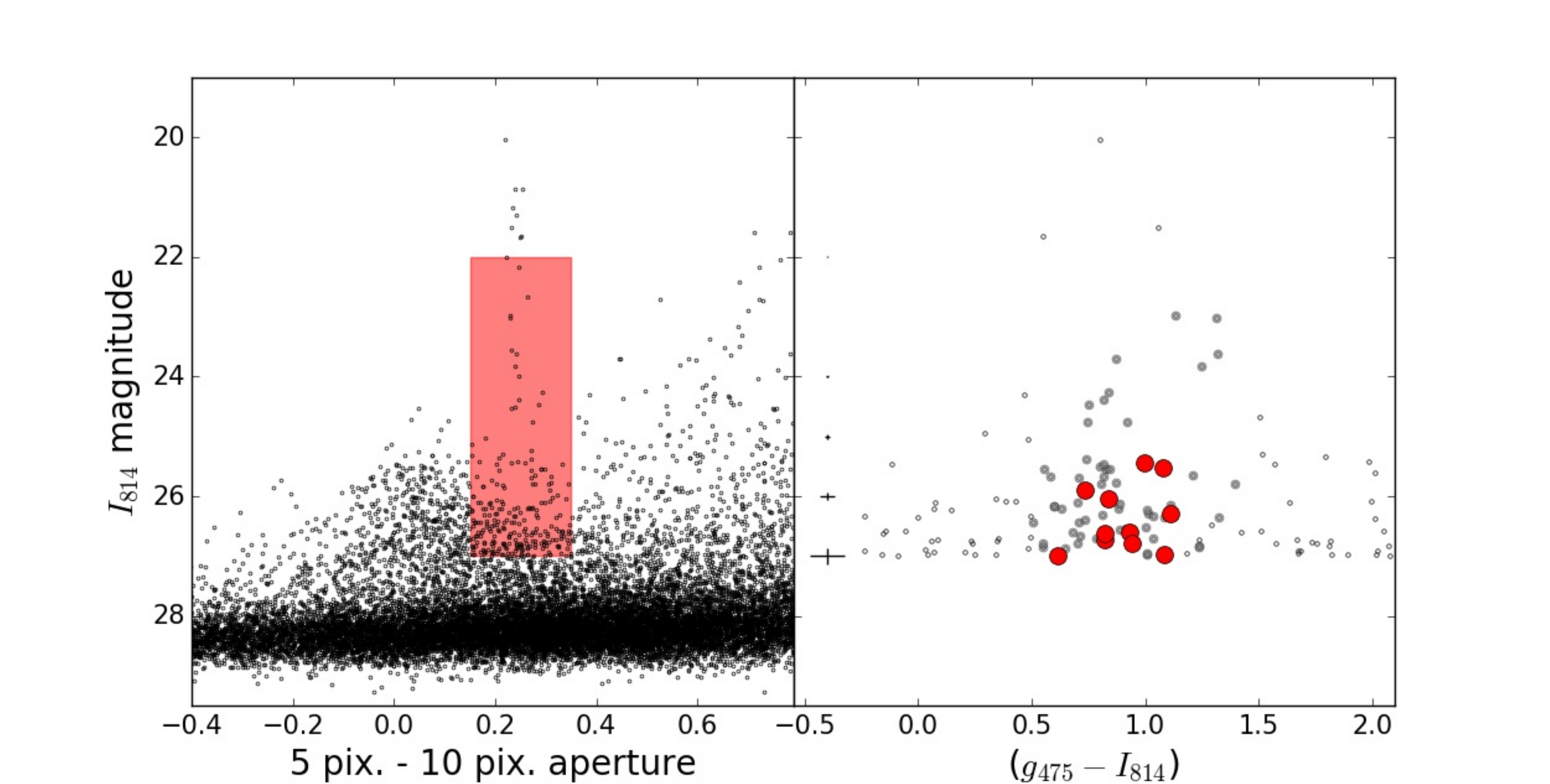}
\caption{Left panel : Selection of point sources in the \I~ imaging.  The locus for point
  sources is located at 0.25. Our point source selection box is shown as the shaded region,
  corresponding to $\pm0.1$ mag about this locus. Right panel: Color-magnitude diagram
  of all points sources brighter than \I = 27.0 (small circles). GC candidates within our color selection
  are shown as filled circles, with GC candidates within 15$''$ radius (two galaxy effective radii)
  from the center of DF17 shown as large red circles. There are 11 GC candidates within this radius. Mean photometric
  uncertainties are indicated to left of the plot.
}\label{fig:Fig2}
\end{figure*}

Our point source selection window is shown in Fig.~\ref{fig:Fig2}. The locus
of point sources lies at 0.25, and we selected objects $\pm0.1$ mag
about this locus. In addition, we imposed a faint magnitude limit
at \I = 26.98, which corresponds to our 5$\sigma$ completeness limit,
and a bright limit at \I = 22.0, which corresponds to $M_I = -13.0$
at our adopted Coma distance, and would include all known MW
GCs. Objects brighter than \I = 22.0 are assumed to be stars.

To narrow our selection further we applied a color cut $0.5<$(\g--\I)$<1.5$ which is consistent with
the expected color ranges for old GCs (e.g., Beasley et al. 2016; Peng et al. 2011).
From our selection criteria, we detect 68 GC candidates.
The color-magnitude diagram of point sources is shown in Fig.~\ref{fig:Fig2}.
Inspection of these sources across the field showed a fairly
uniform distribution, with a clustering of candidates
around DF17. To select potential GC candidates associated with DF17,
we placed a 300-pixel radius aperture about the center of DF17
($\rm{RA}(J2000)=13^{\rm{h}}~01^{\rm{m}}~58.13^{\rm{s}}$,
$\rm{Dec.}~(J2000)=+27^{\rm{o}}~50^{\rm{m}}~11.6^{\rm{s}}$) and
selected all sources within this radius (Fig.~\ref{fig:Fig1}).
A 300 pixel radius corresponds to 15$''$, or $\sim2$ times
the effective radius of DF17 (7$''$; van Dokkum et al. 2015).
This radius is smaller than the extent of the GC system of the
Virgo UDG VCC~1287 (Beasley et al. 2016). We deliberately defined
a smaller region in order to minimize the background contribution
in our GC selection. There are 11 candidates
within this radius and their locations are shown in Figs.~\ref{fig:Fig1}
and ~\ref{fig:Fig2}.

We also measured total magnitudes for DF17 itself by integrating the surface brightness
profile fits. To 10$''$ semimajor axis  we obtain \g=20.17, \V=19.88, \I=19.33.
For Coma, this corresponds to $M_{V_{606}}=-15.12$, or $M_{V,0}=-14.98$
by comparing the ACS \V~ and Johnson $V$ filter responses using an old, metal-poor stellar
population model (Vazdekis et al. 2010) and using the reddening corrections of Schlafly \& Finkbeiner (2011).

To determine the total GC population in DF17, it is necessary
to correct for the contribution from background sources (intracluster GCs and interlopers)
and the limited depth of the  {\it HST/ACS} imaging.
We determine the background contribution by randomly placing 1,000 300 pixel radius apertures across the field
and counting the number of GC candidates within the aperture.
We masked out the region of DF17 itself, and accounted for regions where the selection
region fell outside the detector. Since no other bright galaxies are present in the ACS field,
we masked out no other regions. We find a mean background of
$0.93\pm1.1$ objects within the 300 pixel radius, or $4.7\pm5.6$ objects per arcmin$^2$. 

We determined the total number of GCs in DF17 by assuming that the GC luminosity
function (GCLF) is described by a gaussian function.
Our photometry is 5 $\sigma$ complete to \I = 26.98, corresponding to $M_I=-8.02$ for $m-M=35.0$.
By comparison, Peng et al. (2011) quote a turn-over in the GCLF of $M_{I,AB}=-8.12$
for the Virgo galaxy M87. However, studies of cluster dwarfs suggest
that the turn-over of the GCLF in less massive systems tends to fainter magnitudes, with  $M_{I,AB}=-7.67$
(Miller \& Lotz 2007; Villegas et al. 2010; Peng et al. 2011).

\begin{figure*}[ht]
\epsscale{1.0}
\plotone{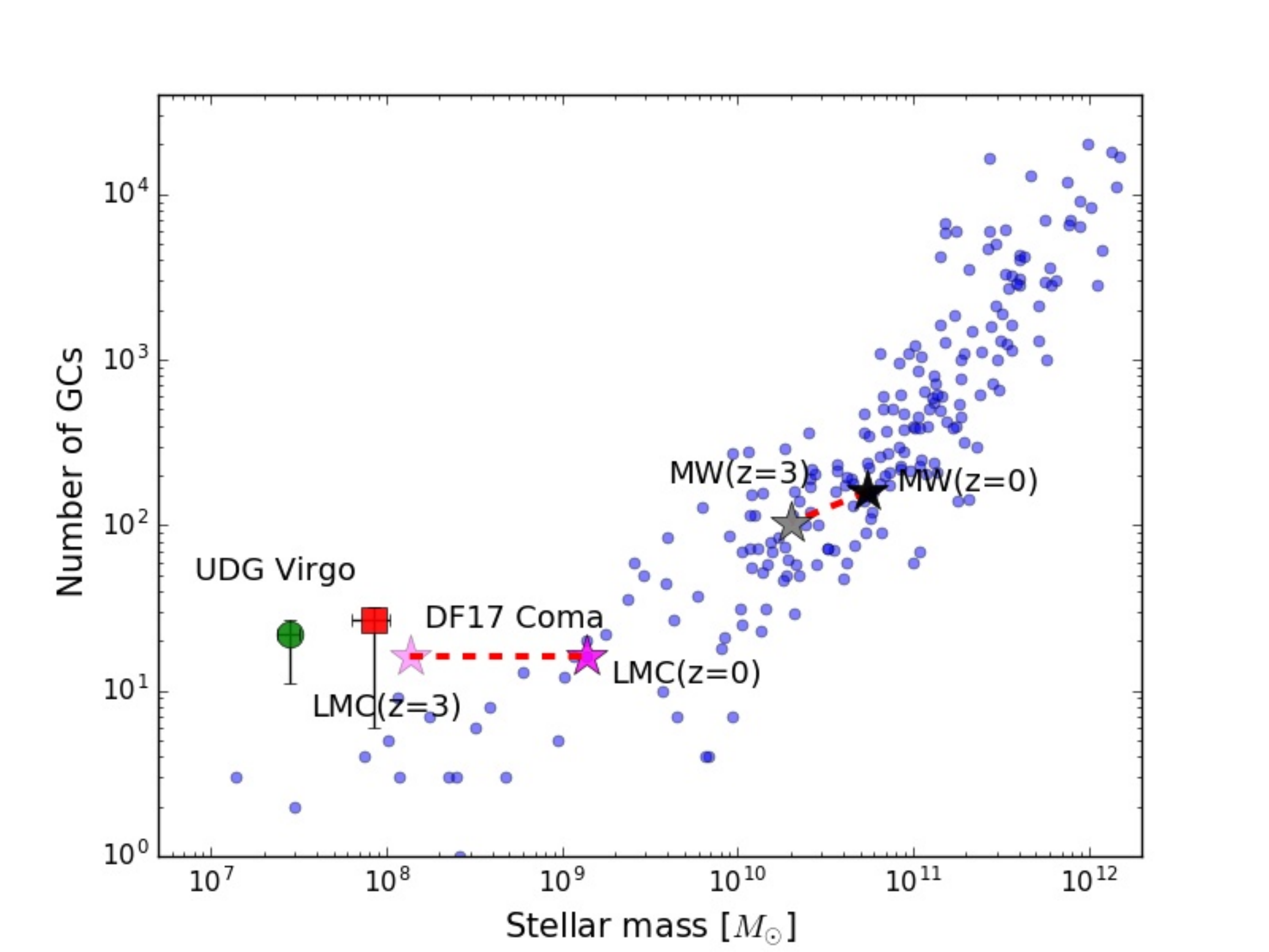}
\caption{Number of GCs versus stellar mass for DF17, VCC~1287 UDG in Virgo (Beasley et al. 2016)
  and nearby galaxies from the Harris et al. (2013) catalog.
  Also shown are the locations of the present-day MW and LMC, and the expected locations
  of these galaxies at redshift 3 assuming a stellar mass evolution for the MW from  Snaith et al. (2014)
  and a GC accreted fraction of 30 percent (Forbes \& Bridges 2010), and a stellar mass evolution
  for the LMC from Leaman et al. (2016; in preparation)
}\label{fig:Fig3}
\end{figure*}

In the case of an ``M87-like'' GCLF, and by integrating a gaussian function, we would detect 54 percent of the
GC population in DF17, implying a total of 19 GCs (including the background correction). In the
``dwarf-like'' case, we would detect the brightest $\sim37.5$ per cent of the GC population, since the GCLF is
both fainter and narrower in these systems, implying a total of 27 GCs. Since our range of estimates is
consistent with that expected for dwarfs, rather than giant galaxies (e.g., Peng et al. 2008), we assume
a dwarf-like GCLF and adopt the latter estimate for the total GC population of DF17, i.e., $N_{\rm GC}=27\pm5$ GCs.
Uncertainties come from the quadrature sum of the poisson and background uncertainties, including a $\pm 3$ Mpc
($\pm 0.07$ mag) uncertainty in the location of the peak of the GCLF. From this we calculate
a $V$-band specific frequency, $S_N=28\pm5$.

\section{Analysis}

\subsection{Comparing total globular cluster populations}

We show in Fig.~\ref{fig:Fig3} the total GC populations as a function of $M_{\rm star}$
for DF17, UDG VCC~1287 in the Virgo cluster (Beasley et al. 2016) and for local galaxies from the catalog
of Harris et al. (2013). Stellar masses are calculated from the $V-$band
luminosities of the systems and stellar mass-to-light ratios from Zibetti et al. (2009).
The lower-limit errorbars for DF17 and VCC~1287 indicate the position of the number of detected
GCs prior to correction for incomplete sampling of the GCLF.
For DF17 we calculate a stellar mass, $M_{\rm star}$=$(8.4\pm2.1)\times10^{7}$\msun, based on the
relation of Taylor et al. (2011) and a conversion from \g-\I~ to Sloan $g$ and $i$ using the Vazdekis et al. (2010)
models.

The locations of the two UDGs indicate that they have significantly poorer GC systems
than the present-day MW
($N_{\rm GC} =160$; $M_{\rm star}~\sim5\times10^{10}$~\msun). Also, the UDGs have
significantly lower stellar mass than the LMC, but have comparable GC populations
(LMC; $N_{\rm GC} =16$; Mackey \& Gilmore 2004).

This is significant because one hypothesis put forward by van Dokkum et al. (2015) as the origin of Coma UDGs
is that they may be quenched $L_*$ systems.
GCs are generally thought to have formed a high redshifts,
as suggested by the ages of MW GCs  ($\geq12$ Gyr, or $z>3$; e.g., VandenBergh et al. 2013)
and the old ages of extragalactic GC systems (Puzia et al. 2005; Strader et al. 2005).
Presumably any quenching of star formation in UDGs took place {\it after} the bulk of the GCs formed,
thereby making the absolute number of GCs an indicator of the nature of UDGs.

\begin{figure*}[ht]
\epsscale{1.2}
\plotone{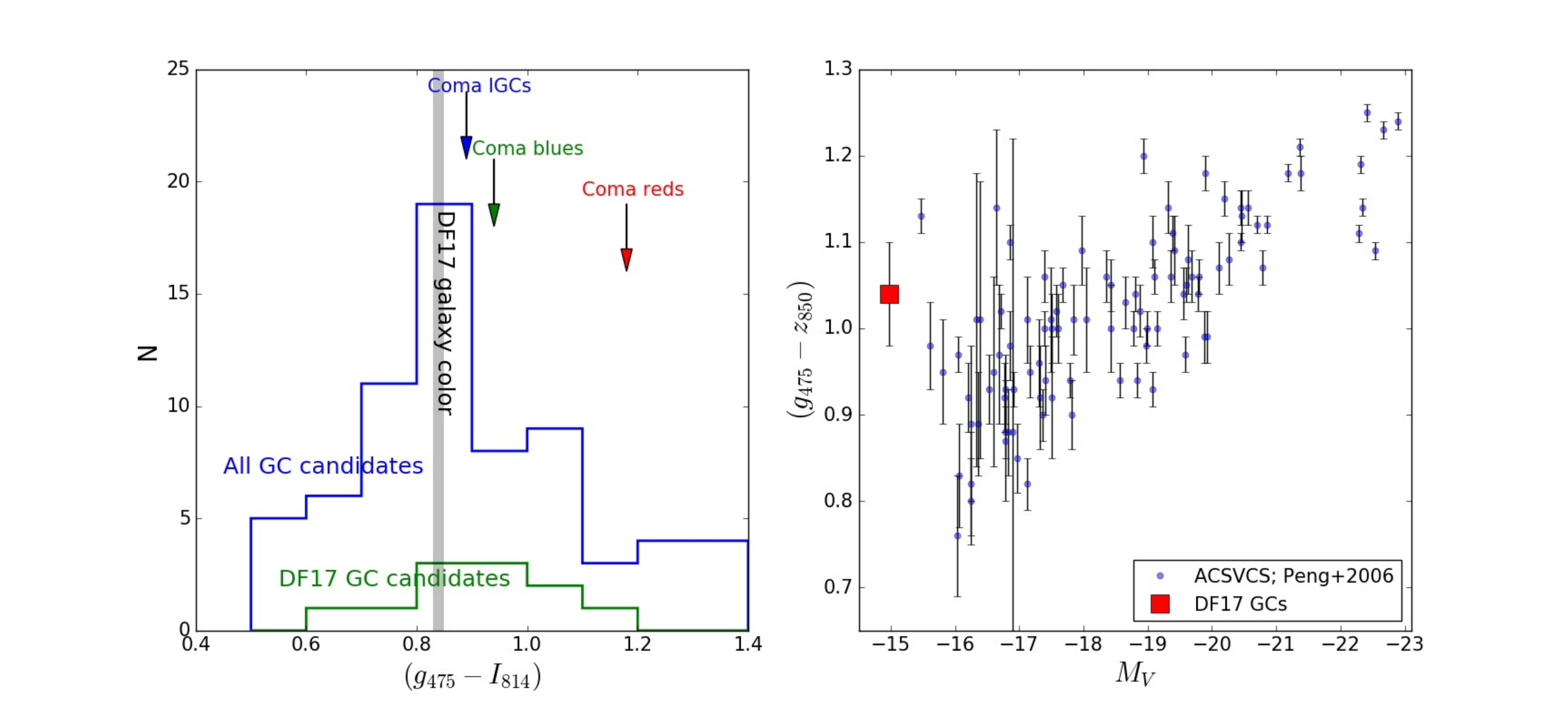}
\caption{Left: Colors of all GC candidates in {\it ACS} field
  compared to that of GC candidates we associate with DF17. Also indicated are
  the locations of the mean colors of intracluster GCs (Coma IGCs), and the blue
  and galaxy subpopulations identified by Peng et al. (2011). The vertical shaded region shows the mean
  color of the galaxy with uncertainties (\g-\I = 0.84$\pm0.1$) within 1 effective radius.
  Right: \g--\z~colors of the DF17 GCs compared to the mean colors from ACSVCS (Peng et al. 2006).
  Our colors have been converted as described in the text. The location of the DF17 GCs is consistent with a
  ``normal'' galaxy with $M_V \sim -19.0$.
}\label{fig:Fig4}
\end{figure*}

In Fig.~\ref{fig:Fig3} we plot the
expected location of the MW at $z=3$, assuming a
stellar mass evolution from Snaith et al. (2015), or a factor of $\sim2.5$ change
in stellar mass over this period. We assume that the
{\it in situ} population of MW GCs formed at high redshift constitutes $\sim70$ per cent
of the present-day GC population, with the remaining $\sim30$ per cent later accreted (Forbes \& Bridges 2010).
The direction of evolution of the MW is indicated by the dashed line
in Fig.~\ref{fig:Fig3}. We find that, assuming a Milky-Way like star formation history,
quenching occurring at $z=3$ does not reproduce the stellar mass or, perhaps more importantly,
the total GC population observed in DF17 (or VCC 1287). Even assuming quenching at $z=5$
(a factor of $\sim5$ change in stellar mass) does not place the MW near
the stellar mass of DF17.

Performing a similar exercise with the LMC is illustrative. Assuming quenching
at $z=3$ implies a mass evolution of a factor of $\sim10$
to the present day (Leaman et al. 2016; in preparation),
placing the LMC (and its GC system) at a stellar mass similar to DF17.

Further, we also obtain an estimate of the virial mass of DF17 by using $N_{\rm GC}$ as a proxy
for total halo mass (Beasley et al. 2016; Harris et al 2013).
For $N_{\rm GC} =27\pm5$ GCs, we obtain $M_{\rm vir} = (9\pm2)\times10^{10}$~\msun, using
the relation of Harris et al. (2013).
The uncertainty on this mass estimate comes purely from the uncertainty in $N_{\rm GC}$,
the scatter in mass in the $N_{\rm GC}-M_{\rm vir}$ relation is at least a factor of two
in the dwarf galaxy regime (Harris et al. 2013).
This yields $M_{\rm vir}$ / $M_{\rm star}\sim1000$.

This virial mass is very similar to that of VCC~1287
$(8.0\pm4.0)\times10^{10}$~\msun; Beasley et al. 2016). Interestingly,
it is also similar to mass estimates for the LMC ($\sim1\times10^{11}$\msun; G\'omez et al. 2015)
- i.e., a relatively massive dwarf galaxy. 

\subsection{Colors of the globular clusters}

The color distribution of DF17 GCs is shown in the left panel of Fig.~\ref{fig:Fig4}, where
we compare with GCs in the field which we regard as intracluster GCs (IGCs).
For the IGCs, we measure a mean (\g--\I) = $0.89\pm0.03$, $\sigma$(\g--\I)=0.21.
This color is identical to that measured by Peng et al. (2011) for IGCs
in the Coma cluster. For the DF17 GCs, we measure a mean (\g--\I) = $0.91\pm0.05$, $\sigma$(\g--\I)=0.15,
identical to the IGCs, and only slightly bluer than the (\g--\I) = 0.94 obtained
by Peng et al. (2011) for the blue GCs these authors associate with massive galaxies.
By contrast, the red peak of the Coma cluster galaxies are substantially redder,
with (\g--\I) = 1.18 (Peng et al. 2011). Thus, based on their colors, both the IGC populations in Coma
and also the outer regions of the massive Coma galaxies may comprise, at least in
part, of GCs from systems similar to DF17.

To compare the colors of the DF17 GCs with a larger sample of galaxies
we turned to the results for the color distributions of GCs from
the ACS Virgo Cluster Survey (ACSVCS; Peng et al. 2006). The ACSVCS
observed 100 galaxies in the \g~and \z~ bands, whereas our observations
include \g~and \I. We searched the literature for a conversion
between \g--\I~and \g--\z~ but found none, therefore we made our own.
We convolved the ACS filter throughput curves for \g, \I~and \z~from
the ACS webpage with the empirically-based model spectra of MIUSCAT (Vazdekis et al. 2012)
selecting 12 Gyr ages for a range of metallicities. The resulting relation between
the predicted  \g--\I~ and \g--\z~ colors is linear, and from linear regression we obtain:

\begin{equation}
  g_{475}-z_{850} = (g_{475}-I_{814})\times 1.023 + 0.128
  \label{eq1}
\end{equation}

and, for completeness,

\begin{equation}
  g_{475}-I_{814} = (g_{475}-z_{850})\times 0.975 - 0.122
\label{eq2}
\end{equation}


Applying Eq.~\ref{eq1} to the mean \g--\I~ colors of the DF17 GCs we
obtain \g--\z$=1.04\pm0.06$.
These are shown in the right panel
of  Fig.~\ref{fig:Fig4}. The mean colors of GC systems of the ACSVCS
galaxies define a color-magnitude sequence; more luminous
galaxies have, on average, redder GC systems. In this context the GCs
of DF17 appear anomalous since they appear too red for the
luminosity of their host galaxy. To lie on the ACSVCS relation,
DF17 would be expected to have $M_V \sim-19.0$, $\sim4$
magnitudes more luminous than its present location.
Typical stellar masses of $M_V \sim -19.0$
galaxies in the ACSVCS are $\sim5\times10^{9}$~\msun. Both
abundance-matching and simulations suggest characteristic virial
masses of these systems to be $\sim1\times10^{11}$~\msun
(e.g., Behroozi et al. 2010; Brook \& Di Cintio 2015; Schaller et al. 2015).
This virial mass estimate from the GC colors is in excellent agreement
with the mass obtained from the GC numbers.

For comparison, stellar and halo mass estimates for the LMC, M33
and the MW are, (LMC: $M_{\rm star}=3\times 10^{9}$~\msun, $M_{\rm vir}=1\times 10^{11}$~\msun;
van der Marel et al. 2014, G\'omez et al 2015;
M33: $M_{\rm star}=6\times 10^{9}$~\msun, $M_{\rm vir}=2 \times 10^{11}$~\msun; Corbelli 2003,
Seigar 2011; MW: $M_{\rm star}=5\times 10^{10}$~\msun, $M_{\rm vir}=1 \times 10^{12}$~\msun;
McMillan 2011; Watkins et al. 2010).

Therefore, both the total numbers and colors of the GCs in DF17 suggest
that they are hosted in a halo more massive than otherwise
suggested by the stellar mass of DF17. Furthermore, the inferred
total mass of the system is characteristic of massive dwarf systems
such as the LMC, rather than the MW.

\section{Discussion and Conclusions}

At least three independent observational tests (Beasley et al. 2016, Rom\'an \& Trujillo 2016 and the one presented here)
 support the idea that UDGs are not failed MW galaxies but instead dwarf galaxies. However, while the stellar masses of
 UDGs are close to 10$^8$ \msun, many of their properties are like those of galaxies a factor of 10
 more massive in stellar mass. For instance, the total number of GCs in DF17 (and in VCC~1287)
 is similar  to that found in the LMC.
 Also, the inferred virial masses of DF17 and VCC~1287 ($M_{vir}\sim 10^{11}$ \msun) locate these objects closer
 to the expectations of a galaxy such as the LMC rather than the MW.
 Moreover, the  sizes of UDGs (1.5$<$r$_e$$<$5 kpc) are not actually uncommon among
 10$^9$~\msun~dwarfs (e.g. Amorisco \& Loeb 2016; Rom\'an \& Trujillo 2016; Shen et al. 2003).
 All these studies support the idea that UDGs are dwarf-like systems,
 and relatively massive dwarfs (in terms of virial mass) - again, perhaps similar to the LMC.

 However, what does seem to make UDGs unusual is their low stellar masses when compared to their
 inferred virial masses (e.g., as shown here and in Beasley et al. 2016). Low stellar masses, but relatively rich GC systems
 suggests that UDGs {\it are} quenched systems,  but whose quenching occured {\it after} the bulk of GC formation occurred.
 Based on the old ages of MW GCs, and present-day stellar masses of UDGs, this possibly places their quenching
 somewhere in the region of $z=3$ (c.f., Fig.~\ref{fig:Fig3}). This quenching redshift is consistent with those seen
 in simulations (Yozin \& Bekki 2015). We view as unlikely the possibility that the low stellar
 masses but rich GC systems of these systems is a result of stripping due to the tidal field of the Coma cluster,
 since simulations suggest that GCs are expected to be lost preferentially over the stars in these systems
 (Smith et al. 2013, 2015).
 
If UDGs are quenched LMC-type galaxies, we should expect that that UDG colors would be compatible with 
the old stellar populations in the LMC. Measuring ages from broad-band colors is fraught with uncertainty,
but the global colors measured for DF17
(\g--\I =0.84) are consistent with larger samples (Rom\'an \& Trujillo 2016), and with a metallicity of
[Fe/H]$\sim-1.0$ at old ages (using the Vazdekis 2010,2015 models).
This metallicity is in agreement with the old ($>10$ Gyr; $z>2$) stars in the disk of the LMC (Carrera et al. 2008).

Therefore, we conclude that UDGs constitute a class of quenched dwarfs - perhaps quenched LMC-like galaxies, whose
quenching took place at $z\sim3$. Further mass measurements of UDGs are important, since it is not presently
clear whether the DM fraction of UDGs (Beasley et al. 2016) are compatible with both proposed models
for their formation (e.g., Amorisco \& Loeb 2016) and also the predicted scatter
in simulations (e.g., Garrison-Kimmel et al. 2016). In addition,
detailed stellar population analyses will help elucidate the precise time and mechanism of
quenching in these systems.

\acknowledgments

We thank Javier Rom\'an for many interesting discussions,
Alejandro Vazdekis for computing models for our color conversions and Ryan Leaman for
use of his mass evolution calculations. We also thank the referee for their
review which improved the paper.
The authors of this paper
acknowledge support from grant AYA2013-48226-C3-1-P  from the Spanish Ministry of  Economy and Competitiveness
(MINECO).





\end{document}